\documentclass[twoside]{dis07}
\usepackage[latin1]{inputenc}
\usepackage[dvips]{graphicx,epsfig,color}
\usepackage{wrapfig,rotating}
\usepackage{amssymb,amsmath,array}

\usepackage{latexsym}
\usepackage{fancybox}
\usepackage{exscale}

\def\be{\begin{equation}}
\def\ee{\end{equation}}

\begin{document}
\title{The ratio of $\sigma_L/\sigma_T$ in DIS at low $x$}

%***********************************************************************
% AUTHORS INFORMATION AREA
%***********************************************************************
\author{Dieter Schildknecht$^{1,2}$ 
%
% Optional short acknowledgment: remove next line if non-needed
%\thanks{This is an optional funding source acknowledgment.}
%
% DO NOT MODIFY THE FOLLOWING '\vspace' ARGUMENT
\vspace{.3cm}\\
%
% Addresses and institutions (remove "1- " in case of a single institution)
1- Fakult{\"a}t f{\"u}r Physik, Universit{\"a}t Bielefeld,
Universit{\"a}tsstrasse 25, D-33615 Bielefeld, Germany
%
% Remove the next three lines in case of a single institution
\vspace{.1cm}\\
2- MPI f{\"u}r Physik, M{\"u}nchen,
F{\"o}hringer Ring 6, D-80805 M{\"u}nchen, Germany\\
}
%***********************************************************************
% END OF AUTHORS INFORMATION AREA
%***********************************************************************

\maketitle

\begin{abstract}
Assuming helicity independence for $q \bar q$ scattering in the color-dipole
picture, or, equivalently proportionality of sea quark and gluon distributions,
we find $R(W^2,Q^2) \cong 0.5$ at large $Q^2$, where $R (W^2, Q^2)$ denotes
the ratio of the longitudinal and transverse photoabsorption cross sections.
The forthcoming direct measurements of $R (W^2, Q^2)$ allow one to test the
underlying hypotheses.
\end{abstract}

%\section{Submission}
\noindent
This is a brief summary of my talk at DIS 2007. We also refer to the slides
of the talk, available under
\verb$http://indico.cern.ch/confAuthorIndex.py?confId=9499$.
It was recently noted \cite{Ewerz} that the dipole picture \cite{Nikolaev}
 of deep 
inelastic scattering
at low $x \cong Q^2/W^2 \le 0.1$,
\be
\sigma_{\gamma^*_{L,T} p} (W^2, Q^2) = \sum_q \, \int d^2 r_\perp 
\omega_{L,T}^{(q)} (Qr_\perp , Q^2 , m^2_q ) \sigma_{(q \bar q )p} 
(r^2_\perp , W^2),
\ee
allows one to derive an upper bound on the ratio of the cross sections 
induced by longitudinal and transverse photons,
\be
R (W^2 , Q^2) = \frac{\sigma_{\gamma^*_L p} (W^2 , Q^2)}{\sigma_{\gamma^*_T p} 
(W^2 , Q^2)} \le \max_{r_\perp , q} \frac{\omega_L^{(q)}(Q r_\perp , Q^2, 
m^2_q)}{\omega_T^{(q)}(Q r_\perp , Q^2, m^2_q)} = 0.37.
\ee
Since the photon fluctuates  into on-shell $q \bar q$ states
\be
\centerline{\epsfig{file=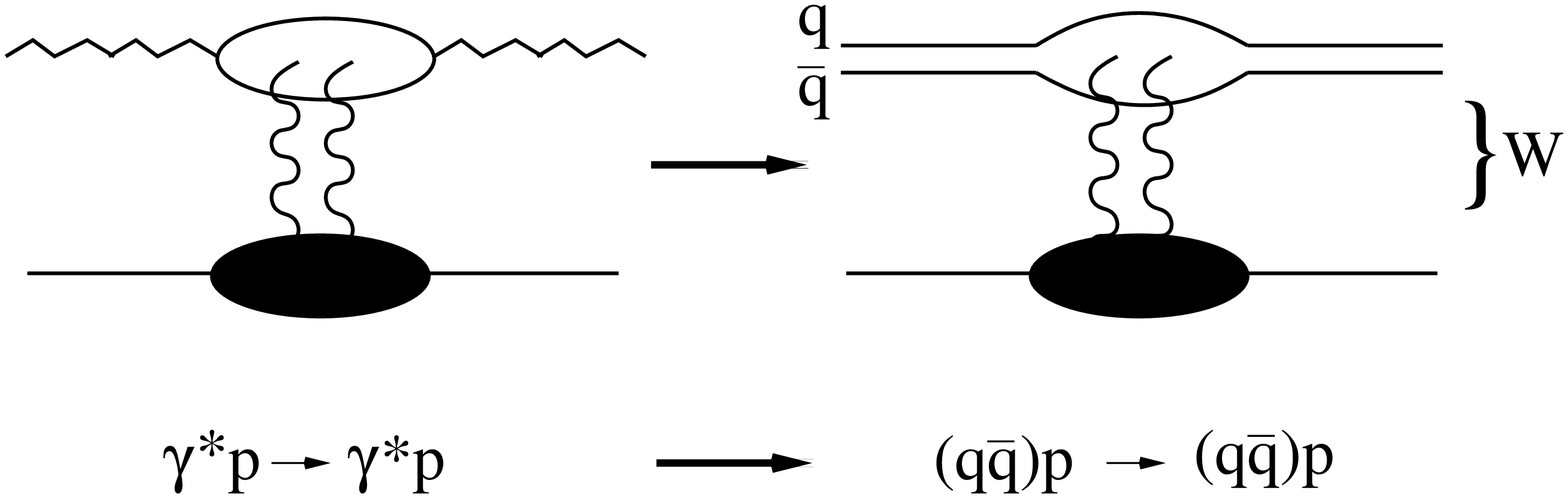,width=12cm}} 
\ee
the $q \bar q$-scattering process entering the virtual
Compton-forward-scattering amplitude is identical to the $q \bar q$-scattering
process of on-shell $q \bar q$ states. Accordingly, as indicated in (1), the
cross section factorizes into a $Q^2$-dependent probability density and a
$W^2$-dependent (rather than $x$-dependent) dipole cross section. The 
relevance of the energy $W$ as the dynamical variable in the low-$x$ 
diffraction region may be traced back to the representation of low-$x$
deep inelastic scattering in terms of generalized vector
dominance \cite{Sakurai} some thirty-five years ago. For the connection between
the dipole picture and generalized vector dominance compare also refs. 4 and 5
and the recent review in 6. The dependence on $W$ rather than $x$ on the
right-hand side in (1) was recently stressed by Ewerz and Nachtmann \cite{Ewerz}
in their very elaborate and explicit treatment of the foundations of the dipole
picture.
\begin{figure}
\vspace*{-5cm}
\epsfig{file=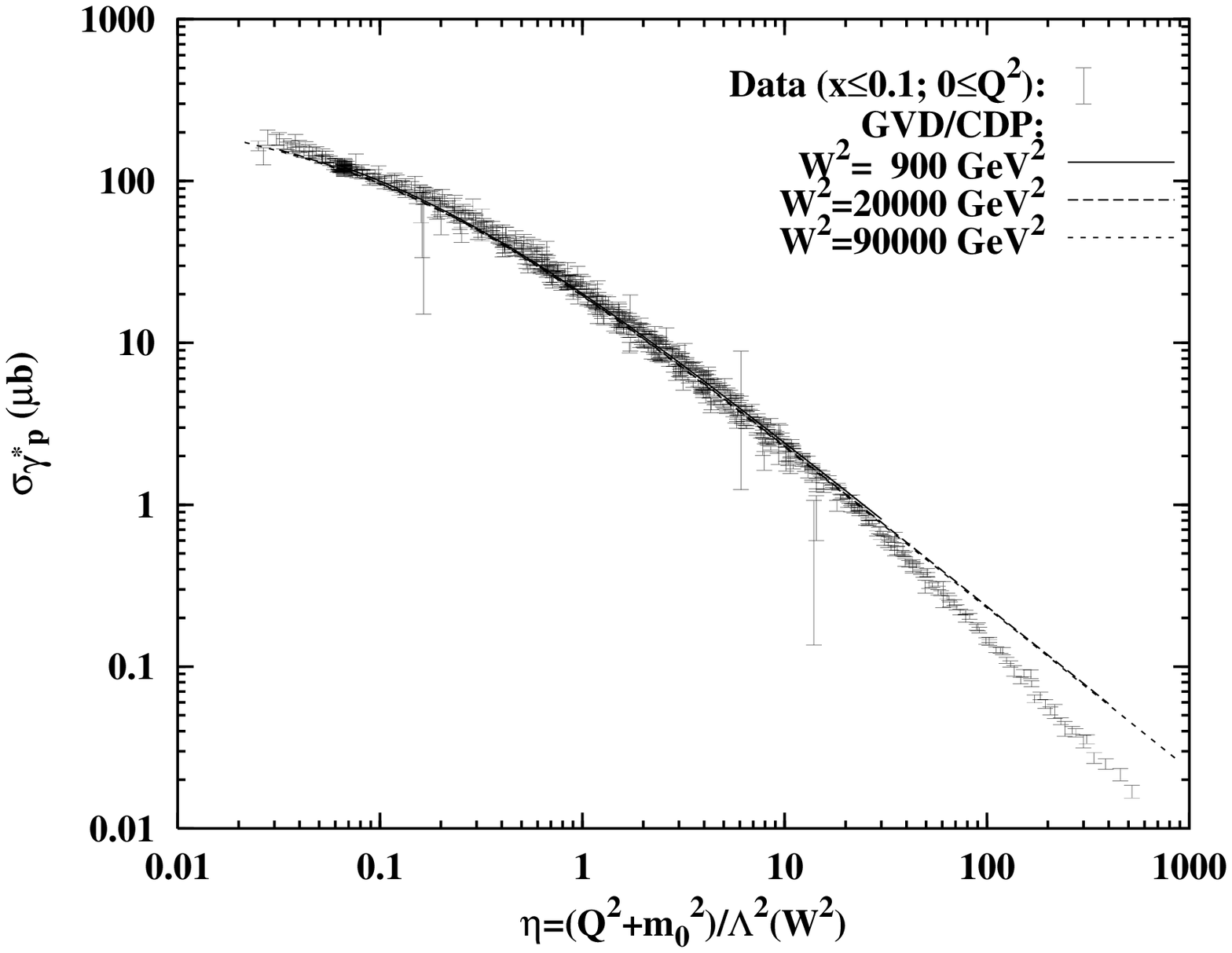,width=6.5cm} 
\hspace{0.3cm}
\epsfig{file=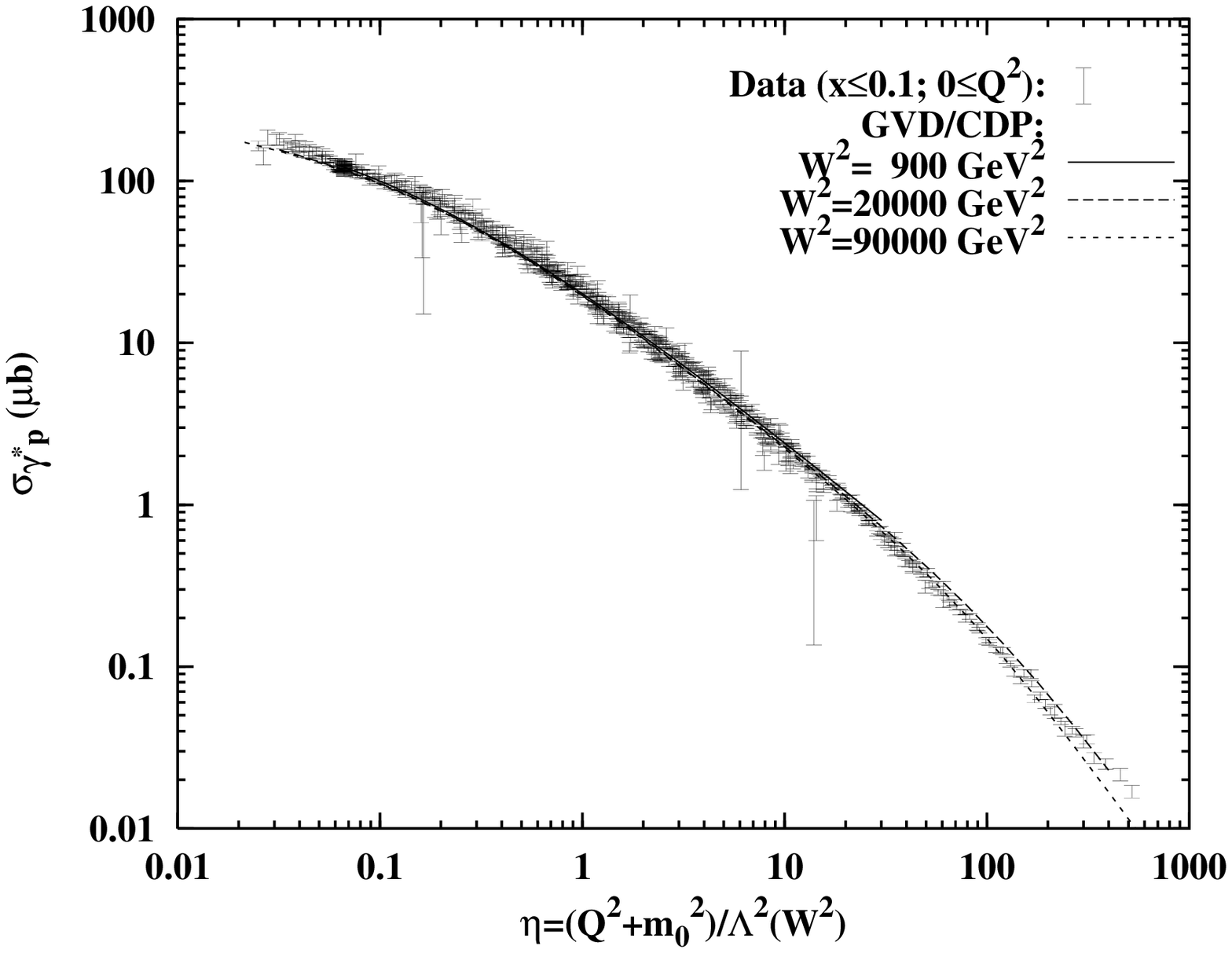,width=6.5cm} 
\caption{The total photoabsorption cross section 
%as a function of $\eta$, 
for $m^2_1=\infty$ and
$m^2_1 = 484 GeV^2$.}
\end{figure}
%\vspace*{-1cm}
\begin{wrapfigure}{l}{0.35\columnwidth}
\centerline{\includegraphics[width=0.35\columnwidth]{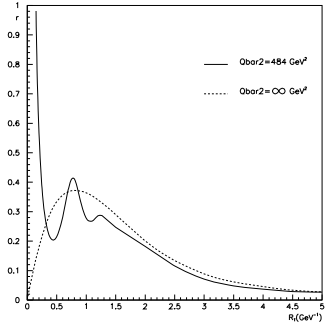}} 
\vspace*{-0.5cm}
\caption{The ratio (7)}
\end{wrapfigure}
Since the available energy, $W$, is finite, the mass, $M_{q \bar q}$, of
the contributing color dipoles must be bounded \cite{Kuroda, Ewerz},
\be
M^2_{q \bar q} = \frac{\vec k^{~2}_\bot}{z (1-z)} \le m_1^2 \equiv 
\bar Q^2 << W^2
\ee
The bound $\bar Q^{~2}$ must be identified with the upper limit of the
diffractively produced masses that is expected to be substantially below
the available energy. In our representation of the HERA data, we used a
value of \cite{Kuroda}
\be
\bar Q^2 \equiv m_1^2 = (22 GeV)^2,
\ee
that was abstracted from the effective upper end of the diffractive
mass spectrum observed at HERA. In Figure 1, we show that the introduction
of the bound \cite{Kuroda} (5) extends the range of validity of the
representation of the cross section in terms of the scaling 
variable \cite{DIFF2000, Surrow}
\be
\eta = \frac{Q^2 + m_0^2}{\Lambda^2_{\rm sat} (W^2)}
\ee
to the region of large values of $\eta$. Since $\bar Q^2$ is determined by
the upper limit of diffractively produced masses, $\bar Q^2$ increases slowly
with increasing energy. To adopt a constant value for the HERA energy range
must be considered as an approximation.

We have analyzed\cite{KuSchi} the effect of the restriction (4,5) on the
ratio $R (W^2, Q^2)$ in (2). The probability density to find a dipole of
size $r_\bot$ in the (virtual) photon now becomes dependent on
$\bar Q^{~2}$. The ratio of the probability densities in (2), for finite
$\bar Q^{~2}$ diverges in the limit of small dipoles, $r_\bot \to 0$,
\be
%\[
r^{(q)} \left( Q r_\perp , \frac{Q^2}{\bar Q^2} \right) = 
\frac{\omega_L^{(q)} (Q r_\perp , Q^2 , \bar Q^2)}{\omega_T^{(q)} 
(Q r_\perp , Q^2 , \bar Q^2 )}
\mathop{\sim}\limits^{r_\perp \to 0} 
 \left\{ \begin{array}{l@{\quad,\quad}l}
Q^2 r^2_\perp \to 0  & {\rm for~~} 
\bar Q^2 \to \infty \\
\frac{1}{r^2_\perp \bar Q^2} \to \infty & {\rm for~~}  \bar Q^2 
{\rm finite}  \end{array}
\right. %\]
\ee
and the bound (2) turns into the trivial statement
\be
0 \le R (W^2, Q^2) < \infty,
\ee
i.e. the derivation of an upper limit for $R(W^2, Q^2)$ fails, once a finite
value for $\bar Q^2$ is adopted.
Compare Figure 2 for the ratio of the probability densities, where for
illustration the value (5) for $\bar Q^2$ is used.

Actually, the representation (1) of the dipole picture must be applied in
conjunction with color transparency\cite{Nikolaev}
\begin{eqnarray}
\sigma_{(q \bar q)p} (r^2_\perp , W^2)& =& \int \, d^2 \vec l_\perp \, \tilde
\sigma_{(q \bar q)p} (\vec l^{~2}_\perp , W^2)(1 - e^{-i\vec l_\perp 
\vec r_\perp})\nonumber \\
% & & \\
& \simeq & \vec r^{~2}_\perp \,\, \frac{\pi}{4} \int \, d \vec l^{~2}_\perp
\vec l^{~2}_\perp  \tilde\sigma_{(q \bar q) p}
(\vec l^{~2}_\perp , W^2 ) , \,\,\,\,\,\,
{\rm for} \, \vec r^{~2}_\perp \to 0.
\end{eqnarray}
Here, $\vec l_\bot$ denotes the transverse momentum of the gluon 
absorbed by the $q \bar q$ pair in the forward-scattering amplitude, where
two gluons of opposite transverse momentum couple to the $q \bar q$ pair.
Since both, transitions $M_{q \bar q} \to M_{q \bar q}$ as well as
$M_{q \bar q} \to M^\prime_{q \bar q}$, occur, the restriction (4) is to
be supplemented by
\be
M^{\prime 2}_{q \bar q} = \frac{(\vec k_\bot + \vec l_\bot)^2}{z (1-z)} 
< \bar Q^2.
\ee
Noting that the momentum of the gluon is entirely independent of the transverse
momentum of the quarks, $k_\bot$, restrictions (4) and (10) together
require
\be
\vec l^{~\prime 2}_\bot = \frac{\vec l^{~2}_\bot}{z (1-z)} << \bar Q^2,
\ee
i.e. the effective change in mass of the $q \bar q$ state by gluon
absorption must be much smaller than the upper bound $\bar Q^2$, where
$Q^2 = m_1^2 \simeq (22 GeV)^2$ at HERA.

In order to investigate the effect of the restrictions (4) and (11) on
$R (W^2, Q^2)$, we appropriately start\cite{KuSchi} with the limit of
$\bar Q^2$ large compared with the effective  value of the gluon transverse
momentum, that is with the limit  of $\bar Q^2 \to \infty$. For $Q^2$ large 
compared with the effective value of $\vec l^{~\prime 2}_\bot$, i.e.
$Q^2 >> <\vec l^{~\prime 2}_\bot>$, where $<\vec l^{~\prime 2}_\bot>$ is
proportional to the ``saturation scale'' $\Lambda^2_{\rm sat} (W^2)$, we find
\be
R (W^2, Q^2) = \underbrace{ \frac{\int d y y^3 K^2_0 (y)}
{\int d yy^3 K^2_1(y)}}_{\frac{1}{2}}
\cdot \frac{\int d \vec l^{~\prime 2} \vec l^{~\prime 2}_\perp 
\bar \sigma_{(q \bar q)^{J=1}_{L} p} (\vec l^{~\prime 2}_\perp , W^2)}
{\int d \vec l^{~\prime 2} \vec l^{~\prime 2}_\perp 
\bar \sigma_{(q \bar q)^{J=1}_{T} p} (\vec l^{~\prime 2}_\perp , W^2)}.
\ee
The ratio of the integrals over modified Bessel functions in (12) yields 
$1/2$. Note that the right-hand side in (12) depends on the ratio of the
$q \bar q$ absorption cross sections for longitudinally and transversely
polarized $(q \bar q)^{J = 1}$ (vector) states. Adopting the assumption of
helicity independence\cite{Surrow,Kuroda}, i.e. equality of the first
moment of the scattering amplitudes for longitudinal and transverse
polarisation, we have from (12)
\be
R (W^2, Q^2) = 0.5.
\ee

We summarize: With color transparency (two gluons coupled to $q \bar q$) 
and the hypothesis of helicity independence, we have $R(W^2, Q^2) = 0.5$
at large $Q^2$. A preliminary investigation\cite{KuSchi} indicates no
substantial change of this result for $\bar Q^2$ finite.

The hypothesis of helicity independence, at large $Q^2$ may be expressed in
terms of a proportionality\cite{SchiKu} of sea quark and gluon distributions.
With the constant of proportionality, $\rho$, we then have 
\be
R(W^2, Q^2 >> \Lambda^2_{\rm sat} (W^2)) = \frac{1}{2 \rho},
\ee
where $\rho = 1$ corresponds to (13). Applying the evolution equation at low
$x$, and large $Q^2$, one finds\cite{SchiKu, Moriond} a correlation between
$\rho$ and the exponent in the $W^2$ dependence of the saturation scale,
\be
\Lambda^2_{\rm sat} (W^2) = const. \left( \frac{W^2}{1 GeV^2} \right)^{C_2}
\ee
given by
\be
(2 \rho + 1) C_2^{theor.} 2^{C_2^{theor.}} = 1.
\ee
Compare Table 1. 

\begin{wraptable}{l}{0.65\columnwidth}
\centerline{\begin{tabular}{|c|c|c|c|c|}
\hline
& & & & \\
$\rho$ & $C_2^{theor.}$ & $\alpha_s \cdot {\rm glue}$ & 
$\sigma_{\gamma^*_L}/\sigma_{\gamma^*_T}$ & $F_2 \left( \frac{Q^2}{x} \right)$
\\ \hline
$\to \infty$ & 0 & $\ll {\rm sea}$ & 0 & $(Q^2/x)^0 = const.$ \\
1 & 0.276
& {$\approx {\rm sea}$}
& {$ \sim \frac{1}{2}$} & {$(Q^2/x)^{0.276}$} \\
0 & 0.65  & $> {\rm sea}$ & $\infty$ & $(Q^2/x)^{0.65}$\\ \hline
\end{tabular}}
\caption{Results for $C_2^{theor.}$ for different values of $\rho$}
\end{wraptable}

The coincidence of the theoretical value of $C_2^{theor.}$ with the 
fit\cite{Surrow} to the experimental data, $C_2^{\rm exp} = 0.27 \pm 0.1$,
supports helicity independence with $\rho = 1$, i.e. $R (W^2, Q^2) = 0.5$
at large $Q^2$. Measurements of $R (W^2, Q^2)$ allow one to directly test
the limits of the assumed proportionality of sea and gluon distributions
that is equivalent to helicity independence and correlated with the rise
of $F_2 (W^2 = Q^2/x)$ as a function of $x$ at fixed $Q^2$.

\vspace*{-0.3cm}

\section{Acknowledgments}

Many thanks to Kuroda-san for collaboration.
Supported by DFG-grant Schi 189/6-1.

\begin{footnotesize}
% IF YOU DO NOT USE BIBTEX, USE THE FOLLOWING SAMPLE SCHEME FOR THE REFERENCES
% ----------------------------------------------------------------------------

% ----------------------------------------------------------------------------

% IF YOU USE BIBTEX,
% - DELETE THE TEXT BETWEEN THE TWO ABOVE DASHED LINES
% - UNCOMMENT THE NEXT TWO LINES AND REPLACE 'Name_Of_Your_BibFile'

%\bibliographystyle{unsrt}
%\bibliography{Name_Of_Your_BibFile}
% example of Name_Of_Your_BibFile.bib
% @Article{Turcato:2006ch,
%      author    = "Turcato, M.",
%  collaboration = "ZEUS and H1",
%      title     = "Lepton flavour violation and charmonium physics at HERA",
%      journal   = "Nucl. Phys. Proc. Suppl.",
%      volume    = "162",
%      year      = "2006", 
%      pages     = "283-287",
%      SLACcitation  = "%%CITATION = NUPHZ,162,283;%%"
% }
% 
% @Unpublished{Gogitidze:2007du,
%      author    = "Gogitidze, N.",
%  collaboration = "H1", 
%      title     = "Prompt photons and particle momentum distributions at
%                   HERA", 
%      year      = "2007",
%      note    = "hep-ex/0701033",
%      SLACcitation  = "%%CITATION = HEP-EX 0701033;%%"
% }

\end{footnotesize}

% ****************************************************************************
% END OF BIBLIOGRAPHY AREA
% ****************************************************************************

\end{document}